\documentclass[aps,prd,twocolumn,showpacs,nofootinbib]{revtex4}

\usepackage{graphicx,amssymb,bm}

\begin{document}

\title{Dark matter annihilation or unresolved astrophysical sources?\\
Anisotropy probe of the origin of cosmic gamma-ray background}

\author{Shin'ichiro Ando}
\email{ando@tapir.caltech.edu}
\affiliation{Theoretical Astrophysics, California Institute of
Technology, Pasadena, CA 91125}
\affiliation{Kellogg Radiation Laboratory, California Institute of
Technology, Pasadena, CA 91125}

\author{Eiichiro Komatsu}
\affiliation{Department of Astronomy, University of Texas at Austin,
Austin, TX 78712}

\author{Takuro Narumoto}
\affiliation{Department of Astronomy, School of Science, Kyoto
University, Kyoto 606-8502, Japan}

\author{Tomonori Totani}
\affiliation{Department of Astronomy, School of Science, Kyoto
University, Kyoto 606-8502, Japan}

\date{December 16, 2006; accepted March 7, 2007}

\begin{abstract}
The origin of the cosmic gamma-ray background (CGB) is a longstanding
 mystery in high-energy astrophysics.
Possible candidates include ordinary astrophysical objects such as
 unresolved blazars, as well as more exotic processes such as dark matter
 annihilation.
While it would be difficult to distinguish them from the mean intensity
 data alone, one can use {\it anisotropy} data instead.
We investigate the CGB
 anisotropy both from unresolved blazars and dark matter annihilation (including
 contributions from dark matter substructures), and find that the
 angular power spectra from these sources are very
 different. 
We then focus on detectability of dark matter annihilation signals
 using the anisotropy data, by treating the unresolved blazar component as a known
 background. 
We find that the dark matter signature should be detectable in the
 angular power spectrum of the CGB from two-year all-sky observations with
 the Gamma Ray Large Area Space Telescope (GLAST), as long as the dark
 matter annihilation contributes to
 a reasonable fraction, e.g., $\agt 0.3$, of the CGB at around 10 GeV.
We conclude that the anisotropy measurement of the CGB with GLAST should
 be a powerful tool for revealing the CGB origin, and potentially for
 the first detection of dark matter annihilation.
\end{abstract}


\maketitle

\section{Introduction}
\label{sec:Introduction}

What is the energy 
budget of the GeV sky?
The origin of the cosmic gamma-ray
background (CGB) in the GeV region, which was discovered by
the Energetic Gamma Ray Experiment Telescope
(EGRET)~\cite{Sreekumar1998,Strong2004a}, is a longstanding mystery.
Unresolved blazars, a beamed population of active galactic nuclei
(AGNs), have been the most popular explanation for the CGB
\cite{Padovani1993,Stecker1993,Salamon1994,Chiang1995,
Stecker1996,Chiang1998,Mucke2000,Narumoto2006,Giommi2006,Dermer2006};
however, even with the latest determination of the gamma-ray luminosity
function (GLF), it has been shown that
only 25--50\% of the CGB can be explained by
unresolved blazars alone~\cite{Narumoto2006}.
Other astrophysical sources include clusters of galaxies, from which gamma rays are
emitted by either hadronic collisions between shock-accelerated protons
and the surrounding medium~\cite{Berezinsky1997,Colafrancesco1998,Pfrommer2003} or the
inverse-Compton scattering of relativistic
electrons off the cosmic microwave background (CMB)
photons~\cite{Loeb2000,Totani2000,Waxman2000,Keshet2003,Miniati2003,Gabici2003a,Gabici2003b,Gabici2004}.
(Note, however, that such gamma rays have not been detected towards
known clusters of galaxies yet~\cite{Reimer2003} though there is some
evidence~\cite{Kawasaki2002,Scharf2002}.)
The CGB may be solely coming from these two unresolved
astrophysical sources at the comparable level for each, or other
sources may be required to explain it.

What could be the other sources of the CGB? 
It is energetically easy to produce gamma-ray photons in the right
energy region by decay or annihilation of heavy (i.e., 10~GeV or
heavier) particles, the particles that have not been discovered yet.
This possibility has been regarded as attractive because (i) more than
80\% of matter in 
the universe is made of nonbaryonic dark matter, 
(ii) theories of supersymmetry predict that the lightest supersymmetric
particles (e.g., neutralinos) are stable and are attractive candidates
for dark matter (in the mass range of GeV to TeV), 
and (iii) these candidate dark matter particles
can annihilate into GeV
gamma rays~\cite{Jungman1996,Bergstrom2000,Bertone2005a}.
The CGB flux from dark matter annihilation in cosmological dark matter halos has
been calculated in the GeV~\cite{Bergstrom2001,Ullio2002,Taylor2003} as well
as in the MeV energies~\cite{AK05a,AK05b}.
Successive works
have shown that this mechanism can account for a large fraction of the CGB
with a reasonable choice of dark matter parameters, with a significant
boost of the signal from substructures~\cite{Oda2005} or
minispikes around intermediate-mass black holes~\cite{Horiuchi2006}
(for the latter, see also Refs.~\cite{Bertone2005b,Bertone2006}).
In addition, with these boosts, we can also avoid an upper limit from
the gamma-ray flux towards the Galactic Center~\cite{Ando2005}. The excess
of the CGB seen at around 3 GeV might be a signature
of dark matter 
annihilation~\cite{Elsaesser2005}.

Ando and Komatsu have recently shown that the angular power spectrum of
{\it anisotropy} in
the CGB may provide a smoking gun discovery of the annihilating dark
matter~\cite{Ando2006a} (hereafter AK06).
In that study the authors focused on the annihilation signal
from dark matter halos with smooth density profiles.
Only semi-quantitative arguments were given for the other
astrophysical sources and the effects of dark matter substructures.
We therefore investigate these two effects in detail in this paper, as
they cannot be ignored if one wants to
discuss whether anisotropy 
can really help
detect the first signature of dark matter annihilation. 

This paper is organized as follows.
In Sec.~\ref{sec:Cosmic gamma-ray background: Mean intensity}, we
explain what the CGB intensity averaged over all the directions looks
like, for both the dark matter annihilation (Sec.~\ref{sub:Dark matter
substructure}) and blazars (Sec.~\ref{sub:Blazars}).
We then turn our attention to the CGB anisotropy from dark matter
substructure and blazars in Secs.~\ref{sec:Cosmic gamma-ray background
anisotropy I: dark matter substructure} and \ref{sec:Cosmic gamma-ray
background anisotropy II: blazars}, respectively, where we present
formulation and results of the angular power spectrum.
Section~\ref{sec:Distinguishing dark matter annihilation and blazars}
is the main part of this paper, devoted to discussion concerning
anticipated anisotropy analysis in the presence of components from both
dark matter annihilation and blazars.
We study the case of other astrophysical sources and discuss the
robustness of our results in Sec.~\ref{sec:Discussion and conclusions},
and also give conclusions in the same section.

\section{Cosmic gamma-ray background: Mean intensity}
\label{sec:Cosmic gamma-ray background: Mean intensity}

In this section we 
calculate the mean intensity (i.e., intensity averaged
over the directions) of the CGB from both dark matter annihilation (with
substructures taken into account)
and blazars, and compare the characteristics of these two components.

\subsection{Dark matter annihilation}
\label{sub:Dark matter substructure}

We include the effect of dark matter substructures as follows:
we assume that substructures consist of a number of subhalos within a
bigger host halo.
These subhalos follow a certain mass function which is still unknown,
but for our purpose we are only interested in quantities that are
averaged over the mass function.
If this mass function is independent of the halo position as we assume,
these {\it averaged} subhalos having the same gamma-ray luminosity would
follow
a smooth density profile of a host halo such as
the one proposed by Navarro, Frenk, and White
(NFW)~\cite{Navarro1996,Navarro1997} with a halo concentration parameter
given in Ref.~\cite{Bullock2001}.
The gamma-ray profile of a halo thus
traces the dark matter density, rather than the density squared which
would be expected if dark matter distribution were smooth~\cite{Ando2006a}.

We define the number intensity, $I_N$, as the number of photons emitted
per unit
area, time, solid angle, and energy range. In a general
cosmological context, it is given by
\begin{equation}
 E I_N (\bm{\hat n},E) = \frac{c}{4\pi}
  \int dz\ \frac{P_\gamma([1+z]E,z,\bm{\hat n}r)}{H(z) (1+z)^4}
  e^{-\tau([1+z]E,z)},
  \label{eq:intensity general}
\end{equation}
where $P_\gamma$ is the volume emissivity (energy of photons per unit
volume, time, and energy range), $H(z)^2 = H_0^2 [(1+z)^3\Omega_m +
\Omega_\Lambda]$ is the Hubble function in a flat universe, and we
assume the standard values for cosmological parameters, $H_0 = 100\ h\
\mathrm{km\ s^{-1}\ Mpc^{-1}}$ with $h = 0.7$, $\Omega_m = 0.3$, and
$\Omega_\Lambda = 0.7$.
We specify a certain direction by a unit vector, $\bm{\hat n}$, position
by a comoving distance vector, $\bm r = r\bm{\hat n}$, and time by a
redshift, $z$ (comoving distance, $r$, is also used interchangeably).
The exponential factor reflects the effect of gamma-ray absorption due
to pair production with the extragalactic background light; such an
effect is negligible in the energy range of interest here.

To evaluate the mean intensity, $\langle I_N(E)
\rangle$, we need
$\langle P_\gamma (E,z) \rangle$.
Let us define the gamma-ray spectrum per subhalo averaged over its mass
function by $\mathcal
N_{\rm sh} (E)$, and the number of these subhalos within a parent halo
of mass $M$
by $\langle N|M\rangle$.
Then, we obtain the mean volume emissivity as
\begin{equation}
 \langle P_\gamma (E,z)\rangle = (1+z)^3 \bar n_{\rm sh} (z)
  E \mathcal N_{\rm sh}(E),
  \label{eq:mean volume emissivity}
\end{equation}
where $\bar n_{\rm sh}(z)$ is the mean comoving number density of subhalos
given by
\begin{equation}
 \bar n_{\rm sh} (z) \equiv \int_{M_{\rm min}}^\infty
  dM\ \frac{dn}{dM}(M,z) \langle N|M \rangle,
  \label{eq:subhalo density}
\end{equation}
and $dn/dM$ is the halo mass function for which we use the expression
given in Ref.~\cite{Sheth1999}.
The function, $\mathcal N_{\rm sh}(E)$, includes all the particle physics
parameters such as the annihilation cross section, $\sigma v$, the 
dark matter mass $m_\chi$, and the gamma-ray spectrum per annihilation $dN_\gamma /
dE$.
For the latter, we use a simple parameterization, i.e., $dN_\gamma /
dE = (0.73/m_\chi) e^{-7.76E/m_\chi} / [(E/m_\chi)^{1.5} + 0.00014]$,
which is a good approximation for supersymmetric neutralino dark matter
particles~\cite{Bergstrom2001}. 
We parameterize the number of subhalos in each parent halo, which is
also known as the Halo Occupation Distribution, as
\begin{equation}
 \langle N|M \rangle =
  \left(\frac{M}{M_0}\right)^\alpha.
  \label{eq:subhalo number}
\end{equation}
If we ignore tidal destruction of subhalos entirely, we obtain $\alpha =
1$, i.e., the number of subhalos is simply proportional to the mass of
the parent halo.
The tidal destruction should also change the gamma-ray emission
profiles in the parent halo, as it works more strongly at inner halo
regions.
However, we adopt the NFW profile for all of our calculations, because
the profile change should exert only secondary effect to our conclusions
as discussed in Sec.~\ref{sub:Results 1}.

\begin{figure}
\begin{center}
\includegraphics[width=8.5cm]{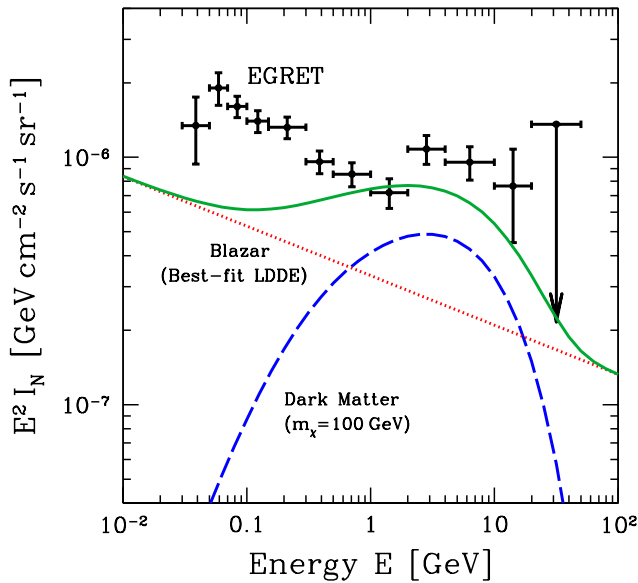}
\caption{The CGB spectrum from dark matter annihilation (dashed) and
 blazars with the best-fit LDDE GLF (dotted). Total intensity is shown
 by the solid curve, and the data points are from the EGRET
 data~\cite{Strong2004a}. \label{fig:spectrum}} 
\end{center}
\end{figure}

Using Eqs.~(\ref{eq:intensity general}) and (\ref{eq:mean volume
emissivity}), we obtain 
the mean intensity as
follows:
\begin{equation}
 \langle I_N(E)\rangle = \int dr\ W([1+z]E,z),
  \label{eq:mean intensity: simple form}
\end{equation}
where 
\begin{equation}
 W(E,z) = \frac{1}{4\pi} \bar n_{\rm sh}(z) \mathcal N_{\rm sh} (E,z)
  e^{-\tau(E,z)}.
  \label{eq:window function}
\end{equation}
Figure~\ref{fig:spectrum} shows the CGB spectrum from dark matter
annihilation, where the particle mass is assumed to be $m_\chi = 100$
GeV.
We do not give a specific value of $\sigma v$ or $M_0$. 
These parameters
are degenerate, but they do not affect predictions of the angular power
spectrum, as we see below.
All we require here is that the predicted intensity becomes comparable
to the observed CGB, 
and this can be done by adjusting these two parameters.
Previous work which included dark matter substructures~\cite{Oda2005}
has shown that 
this is indeed possible with a standard value of the annihilation cross
section, $\sigma v = 3\times 10^{-26}$ cm$^{3}$ s$^{-1}$,
which gives the right amount of the dark matter density in the universe 
if dark matter was thermally produced
in the early universe~\cite{Jungman1996,Bertone2005a}.
On the other hand, anisotropy depends only on $m_\chi$ and
$\alpha$.
We shall therefore vary $\alpha$ and see how the results depend on
$\alpha$,
while we fix the mass at 100 GeV throughout the paper.

\subsection{Blazars}
\label{sub:Blazars}


If a non-negligible fraction of the CGB flux comes from 
astrophysical sources such as blazars and clusters of galaxies, they
inevitably give a {\it 
background} (noise) for the dark matter detection in the anisotropy signature.
It is thus very important to evaluate the contribution from
the unresolved point sources. 
We concentrate on blazars as an
example.

To calculate the mean CGB intensity from blazars one needs 
the GLF of blazars.
We use the latest luminosity dependent density evolution
(LDDE) model, which 
reproduces the observed GLF of the EGRET blazars
better than a traditionally used pure luminosity evolution
model~\cite{Narumoto2006}.
As the LDDE GLF was originally given for the luminosity at 100 MeV, we
need to generalize it to the other energies.
We do this by specifying the spectral shape; here we assume it
to be a power law with a spectral index of $\alpha_\gamma =
2.2$~\cite{Sreekumar1998}.
Then, the luminosity per unit energy range, $\mathcal L$, is connected to
the luminosity, $L_\gamma (100\ \mathrm{MeV})$ ($=E \mathcal L$ at 100~MeV)
adopted in the previous
GLF via the following simple relation:
\begin{equation}
 \mathcal L(E_{\rm em}) =
  \left(\frac{E_{\rm em}}{\mathrm{100\ MeV}}\right)^{1-\alpha_\gamma}
  \frac{L_\gamma}{\mathrm{100\ MeV}},
  \label{eq:luminosity relation}
\end{equation}
The GLF is accordingly replaced with the one defined as the comoving
number density per unit range in $\mathcal{L}$, $\Phi_E
(\mathcal{L},z)$, which is related to the original one through
\begin{equation}
 d\mathcal L\ \Phi_E (\mathcal L,z)
  = dL_\gamma\ \rho_\gamma(L_\gamma,z),
  \label{eq:GLF relation}
\end{equation}
where we show the energy dependence of the new GLF explicitly by
attaching subscript $E$.
Note that $\rho_\gamma$ on the right hand side is given by
Eqs.~(8) and (10) of Ref.~\cite{Ando2006b}.
Using Eqs.~(\ref{eq:luminosity relation}) and (\ref{eq:GLF
relation}), we can rewrite the luminosity and the GLF at any energies as
long as the spectrum is kept to be a power law with the same index.

The photon flux from the source with luminosity $\mathcal L$ at redshift
$z$ at energy $E$ is given by
\begin{equation}
 \mathcal F_E(\mathcal L,z)
  = \frac{(1+z)\mathcal L[(1+z)E,z]}{4\pi d_L^2(z)},
  \label{eq:point source flux}
\end{equation}
where $d_L(z)$ is the luminosity distance out to a source at $z$.
The flux sensitivity for point sources of the EGRET is $F_{\gamma,{\rm
lim}} \simeq 10^{-7}$ cm$^{-2}$ s$^{-1}$ above 100
MeV~\cite{Thompson1993}, and all the unresolved sources that give a flux
{\it below} this threshold contribute to the CGB.
The conversion from the differential flux per energy, $\mathcal F_E$, to the
integrated flux, $F_\gamma$, can easily be performed by integrating over
energy above 100 MeV and assuming the spectrum to be a power law with an
index $\alpha_\gamma$. 
One obtains 
\begin{equation}
 \mathcal F_E = (\alpha_\gamma-1)
  \left(\frac{E}{\mathrm{100\ MeV}}\right)^{1-\alpha_\gamma}
  F_\gamma.
  \label{eq:flux relation}
\end{equation}
We use this equation and Eq.~(\ref{eq:point source flux}) to calculate the limiting source luminosity,
$\mathcal L(\mathcal F_{E,{\rm lim}},z)$, from 
$F_{\gamma,{\rm lim}}$.

We calculate the mean CGB intensity coming from unresolved blazars whose gamma-ray
flux is below $\mathcal F_{E,{\rm lim}}$ from
\begin{eqnarray}
 E \langle I_N(E) \rangle &=& \int_0^{z_{\rm max}}dz\
  \frac{d^2V}{dz d\Omega}
  \int_0^{\mathcal L(\mathcal F_{E,{\rm lim}},z)}
  d\mathcal L\ \Phi_E(\mathcal L,z)
  \nonumber\\&&{}\times
  \mathcal F_E(\mathcal L,z),
  \label{eq:CGB}
\end{eqnarray}
where we use $z_{\rm max} = 5$, and $d^2V/dzd\Omega$ is the comoving
volume per unit redshift and unit solid angle ranges.
We show in Fig.~\ref{fig:spectrum} the CGB spectrum calculated with the
best-fitting LDDE GLF together with the EGRET data.
The predictions fall below the EGRET data, 
accounting for only 25--50\% of the observed CGB
\cite{Narumoto2006,Ando2006b}.

This is presumably either because there is another class of objects which
can contribute to the CGB by equally significant amount, or because the
best-fitting LDDE model is underestimating the true GLF.\footnote{This
may instead be because of underestimated contamination of the Galactic
foreground emission~\cite{Keshet2004a}.}
For the former case, the particle acceleration in other astrophysical
objects may also give a power-law CGB spectrum similar to that of
blazars, and depending on its luminosity function, an unaccounted
fraction of the CGB flux might be attributed to this population.
Since the EGRET data may be explained by a power law component, such
additional sources, plus blazars, 
might give almost full account of the CGB in the GeV region.
One such candidate is galaxy clusters, in which protons or electrons
are accelerated to relativistic energies by shock waves, and the GeV
gamma rays are emitted by either pion production due to the
proton-proton interactions~\cite{Berezinsky1997,Colafrancesco1998,Pfrommer2003} or
inverse-Compton scattering of relativistic
electrons off CMB photons~\cite{Loeb2000,Totani2000,Waxman2000,Keshet2003,Miniati2003,Gabici2003a,Gabici2003b,Gabici2004}.
The latter possibility is also possible, although the GLF parameters
that can account for 100\% of the EGRET data are inconsistent with the
X-ray data at the 4.4-$\sigma$ level. (See Sec.~2.2 of Ref.~\cite{Ando2006b}
for details.)

GLAST is expected to 
detect $1000$--10000 blazars
as point sources \cite{Narumoto2006,Ando2006b}. 
We can thus improve on accuracy of the GLF significantly after GLAST
with much better statistics.

\section{Cosmic gamma-ray background anisotropy I: dark matter
 annihilation}
\label{sec:Cosmic gamma-ray background anisotropy I: dark matter
substructure}

The angular power spectrum of CGB anisotropy calculated by
AK06 does not take into account the effect of dark matter
substructure. In this section we extend their work by including
substructures explicitly.

\subsection{Formulation}
\label{sub:Formulation 1}

The angular power spectrum, $C_l$, is given by
\begin{eqnarray}
 C_l &=& \langle |a_{lm}|^2 \rangle,
  \label{eq:C_l}\\
 a_{lm} &=& \int d\Omega_{\bm{\hat n}}\
  \frac{I_N(E,\bm{\hat n})-\langle I_N(E)\rangle}{\langle I_N(E)\rangle}
  Y_{lm}^\ast(\bm{\hat n}).
  \label{eq:a_lm}
\end{eqnarray}
It is related to the spatial power spectrum of subhalos, $P_{\rm
sh}(k,z)$, through 
\begin{equation}
 \langle I_N(E) \rangle^2 C_l
  = \int \frac{dr}{r^2}\left\{W([1+z]E,z)\right\}^2
  P_{\rm sh}\left(\frac{l}{r},z\right),
  \label{eq:Limber's equation}
\end{equation}
where $W(E,z)$ is again given by Eq.~(\ref{eq:window function}), and
$P_{\rm sh}(k,z)$ may be divided into 1-halo
($1h$) and 2-halo ($2h$) terms:
\begin{eqnarray}
 P_{\rm sh}(k) &=& P_{\rm sh}^{1h}(k) + P_{\rm sh}^{2h}(k),
  \label{eq:spatial power spectrum}\\
 P_{\rm sh}^{1h}(k) &=& \int_{M_{\rm min}}^\infty
  dM\ \frac{dn}{dM}
  \left(\frac{\langle N|M \rangle}{\bar n_{\rm sh}}\right)^2
  |u(k|M)|^2,\nonumber\\
  \label{eq:1-halo term}\\
 P_{\rm sh}^{2h}(k) &=& \left[\int_{M_{\rm min}}^\infty
  dM\ \frac{dn}{dM}
  \frac{\langle N|M \rangle}{\bar n_{\rm sh}}
  b(M) u(k|M)\right]^2
 \nonumber\\&&{}\times
 P_{\rm lin}(k).
 \label{eq:2-halo term}
\end{eqnarray}
Each term means that we correlate two different points in one identical
halo ($1h$) or two distinct halos ($2h$).
Correspondingly, we also have 1-halo and 2-halo terms for the angular
power spectrum, i.e., $C_l = C_l^{1h} + C_l^{2h}$.

The 1-halo term is determined basically by the
density profile of parent halos, $\rho (r|M)$, as we assume that the subhalos
in a parent halo distribute by following $\rho (r|M)$. 
Here, $u(k|M)$ is the Fourier transform of 
$\rho(r|M)/M$.
The 2-halo term, on the other hand, is proportional to the linear matter
power spectrum, $P_{\rm lin}(k)$~\cite{Eisenstein1999}, and depends
on the halo bias, $b(M)$~\cite{Mo1996}. 
Detailed derivations of these results are given in Appendix~\ref{apsub:Dark matter
substructure}.
We calculate the angular power spectrum from Eqs.~(\ref{eq:Limber's
equation})--(\ref{eq:2-halo term}) with the NFW density profile of dark
matter halos.

\subsection{Results}
\label{sub:Results 1}

\begin{figure*}
\begin{center}
\includegraphics[width=12cm]{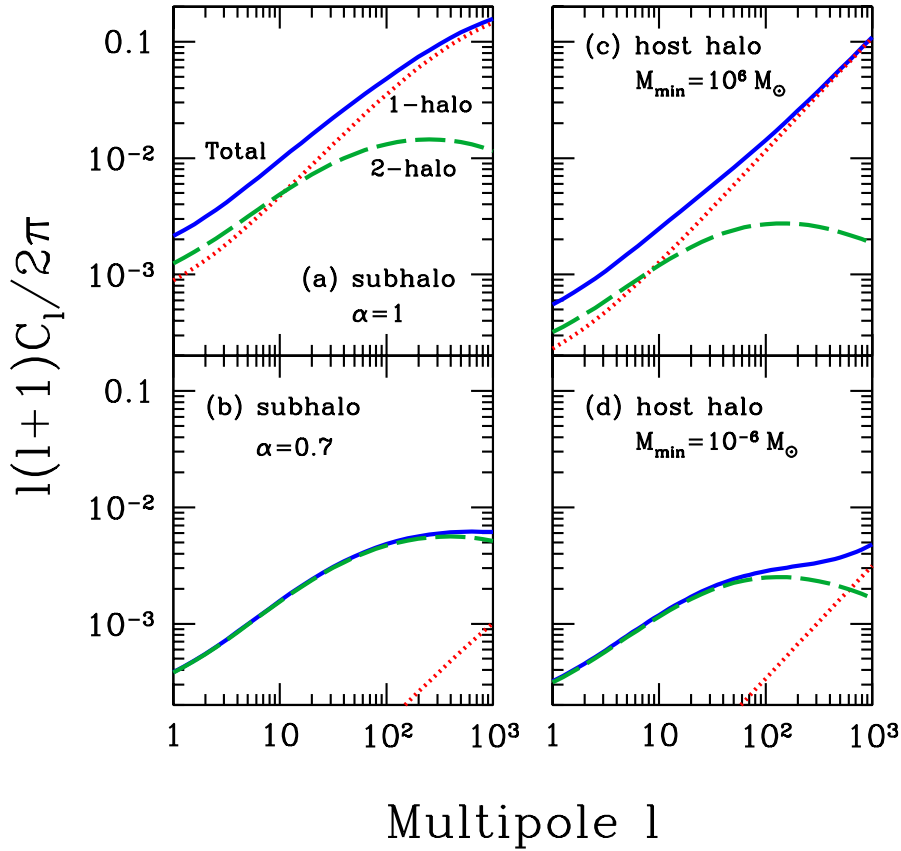}
\caption{{\it Left:} Angular power spectrum, $l (l+1) C_l / 2\pi$, of
 CGB from dark matter annihilation with substructures, 
 (a) $\alpha = 1$ and (b) 0.7. The
 dotted and dashed lines show the contribution from the 1-halo and 2-halo
 terms, respectively, and the solid line is the total. {\it Right:} The same
 as left panels but with smooth dark matter distribution (i.e., no
 substructure; AK06), with the
 minimum mass of (c) $10^6 M_\odot$ and (d) $10^{-6} M_\odot$.
 \label{fig:C_l_dm}}
\end{center}
\end{figure*}

Left panels of Fig.~\ref{fig:C_l_dm} show the results for two
different Halo Occupation Distribution, $\alpha = 1$
(top) and $\alpha = 0.7$ (bottom) [see Eq.~(\ref{eq:subhalo number})].
We have used $M_{\rm min} = 10^6 M_\odot$ in Eqs.~(\ref{eq:1-halo term}) and
(\ref{eq:2-halo term}). The dependence on $M_{\rm min}$ is
very weak, as the CGB flux from annihilation is dominated by massive
halos which host many subhalos. 

The 1-halo term dominates at larger 
$l$'s, or smaller angular scales, as expected. 
We find that the 1-halo term strongly depends on $\alpha$:
the smaller $\alpha$, the larger the contribution from less massive
halos.
As smaller halos are dimmer, one needs to increase the source number
in order to explain the observed CGB flux.
Therefore, this makes the CGB more isotropic, resulting in a reduction of
the 1-halo term for $\alpha = 0.7$.
On the other hand, dependence of the 2-halo term on $\alpha$ is much
weaker as the 2-halo term is essentially given by 
the linear matter power spectrum, $P_{\rm lin}(k)$, times the average
halo bias.
This weak dependence comes from the bias factor that is in the
integrand of Eq.~(\ref{eq:2-halo term}).
If we increase the contribution of low-mass halos by reducing $\alpha$,
then the 2-halo term decreases as such low-mass halos are less biased.

We use these two cases, $\alpha = 1$ and 0.7, to bracket the uncertainty
in $\alpha$.
With $\alpha = 1$, we assume that all the substructures survive the
tidal disruption, which is probably a very optimistic approximation.
For $\alpha = 0.7$, on the other hand, we allow some amount of
disruption effect, and the angular power spectrum is totally dominated
by the 2-halo term, another extreme situation.
For even smaller values of $\alpha$, the 1-halo term disappears but the
2-halo term decreases only slightly; thus we do not consider these cases
any further. 

We also consider the case in which the smooth host halo component
dominates the annihilation signal. (I.e., substructures are unimportant.)
The gamma-ray emission profile is proportional to the
density squared, and this case was in fact carefully studied in AK06.
(This case, however, requires additional assumption such that the 
density profile of the Milky Way is much shallower than NFW, in order
not to violate the constraint 
from the Galactic Center~\cite{Ando2005}.)
The right panels of Fig.~\ref{fig:C_l_dm} show the angular power
spectrum of smooth-halo cases with different values of the minimum mass,
$M_{\rm min}$.
Unlike the subhalo-dominated case, the CGB flux is very sensitive to the
choice of $M_{\rm min}$.
This is because the halo concentration is larger for smaller
halos~\cite{Bullock2001}, and the gamma-ray luminosity from each halo is
very sensitive to the concentration parameter.
We again adopt two extreme cases here, one dominated by 1-halo term and
the other by 2-halo term.
The shape of the 1-halo term on the right panels is steeper than that on
the left panels (subhalo-dominated case), as the signal is more
concentrated at the central region.
We see below that this tendency converges to $C_l^{1h} =
\mathrm{const.}$ when the source is infinitely small as expected for 
point-like astrophysical sources such as blazars.
As for the 2-halo term, the shape is almost the same as the
subhalo-dominated case.
The normalization, however, is smaller, because the contribution from
less massive halos is enhanced when the dark matter distribution is smooth
(owing to large concentration of small halos), and less massive halos
are less biased.

As a final remark in this section, we discuss the case with other
density profiles.
Although the NFW profile is most widely used in the literature, our
knowledge of the density profile is still far from converged.
In particular towards the central region, density might keep rising or
converge at some constant value.
Therefore, one might question about the impact of the different profiles
on the angular power spectrum.
We have already seen the tendency such that the steeper the gamma-ray
emission profile becomes, the harder the 1-halo term of angular power
spectrum gets.
Exactly the same argument applies here as well.
If the profile has a flat core within some radius, the 1-halo term
should be flattened above some $l$ that corresponds to the core radius,
but this modification at such a small scale will not be detected with
GLAST (see discussions about detectability in
Sec.~\ref{sub:Detectability of the dark matter component}).
The flattened profile could also be caused by the tidal disruption of
subhalos.
This means that it controls both the shape (via the profile) and
normalization (via the number distribution, or $\alpha$) of the 1-halo
term; but the latter would be much more prominent.
On the other hand, the 2-halo term would stay almost the same even if we
changed the density profile, thus providing the {\it guaranteed} power
spectrum that is independent of the density profile adopted.

\section{Cosmic gamma-ray background anisotropy II: blazars}
\label{sec:Cosmic gamma-ray background anisotropy II: blazars}

%
%


\subsection{Formulation}
\label{sub:Formulation 2}

When sources are point-like just as blazars, the angular power spectrum
of the CGB is given by
\begin{equation}
  C_l = C_l^{P} + C_l^{C},
\end{equation}
where the first, Poisson term $C_l^{P}$, corresponds to the 1-halo term,
and the second, correlation term $C_l^{C}$, to the 2-halo term. 
The Poisson term represents the shot noise that does not depend on the
multipole $l$'s, while the correlation term is due to the intrinsic
spatial correlation of sources.
These two terms are related to the spatial power spectrum through
\begin{eqnarray}
 C_l^{P} &=& \frac{1}{E^2\langle I_N(E)\rangle^2}
 \int dz\ \frac{d^2V}{dzd\Omega}\
 \nonumber\\&&{}\times
 \int d\mathcal L\ \Phi_E(\mathcal L,z)
 \mathcal F_E(\mathcal L,z)^2,
 \label{eq:pois}\\
 C_l^{C} &=& \frac{1}{E^2\langle I_N(E)\rangle^2}
  \int dz\ \frac{d^2V}{dzd\Omega}P_{\rm
 lin}\left(\frac{l}{r(z)},z\right)
  \nonumber\\&&{}\times
  \left[
   \int d\mathcal L\
   \Phi_E(\mathcal L,z) b_B(\mathcal L,z)\mathcal F_E(\mathcal L,z)
  \right]^2,\nonumber\\
\label{eq:corr}
\end{eqnarray}
where the lower and upper bounds of integration over $z$ and $\mathcal L$
are the same as in Eq.~(\ref{eq:CGB}).
Detailed derivations are given in Appendix~\ref{apsub:Blazars}.
Here the power spectrum of blazars is approximated as $P_B(k;\mathcal
L_1, \mathcal L_2) \approx b_B(\mathcal L_1) b_B(\mathcal L_2) P_{\rm
lin}(k)$, and the blazar bias, $b_B$, represents how strongly blazars
cluster compared with dark matter. (See also Sec.~2.1 of Ref.~\cite{Ando2006b}.)

While the blazar bias, $b_B(\mathcal L,z)$, is currently unknown,
it will probably be measured directly from GLAST blazar
catalog~\cite{Ando2006b}. 
(This measurement is, however, limited to the bias of {\it resolved}
blazars, which can be different from the bias of {\it unresolved} blazars
which contribute to the CGB anisotropy.)
At the moment, one may estimate $b_B(\mathcal L,z)$ from several
approaches including 
the angular and spatial correlation analysis of optical
quasars~\cite{Croom2005,Myers2006} and X-ray selected
AGNs~\cite{Yang2003,Basilakos2005,Gandhi2006}.
These results, however, are not consistent with each other, potentially
due to some observations being biased by a limited field of
view covered, or because there is something wrong in our understanding of the
unified picture of the AGNs.
In any case,  a very wide range of the blazar bias is still
allowed, $b_B \alt 5$; see Sec.~3.2 of Ref.~\cite{Ando2006b} for a
more detailed discussion.

\subsection{Results}
\label{sub:Results 2}

\begin{figure}
\begin{center}
\includegraphics[width=8.5cm]{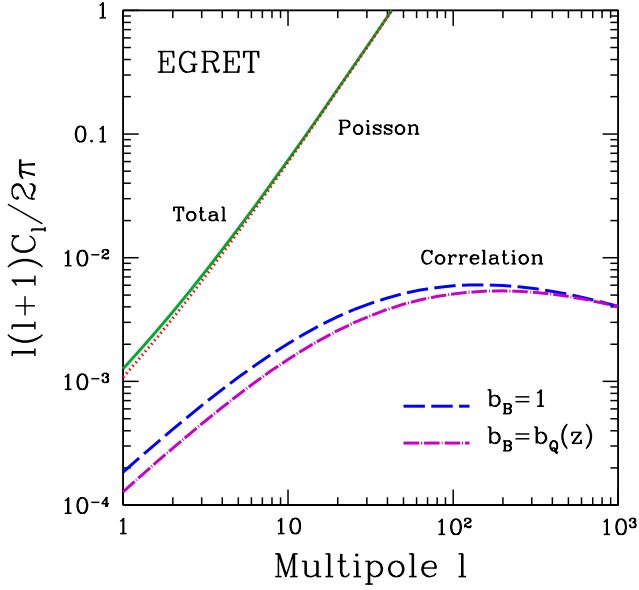}
\caption{Angular power spectrum of the CGB from unresolved blazars expected from the
 EGRET data. Contributions from Poisson term, $C_l^P$, and the
 correlation term, 
 $C_l^C$ with $b_B=1$ ($b_B = b_Q(z)$), are shown by the dotted and dashed
 (dot-dashed) curves, respectively. The total contribution  is shown as
 the solid curve for $b_B = 1$ \label{fig:C_l_EGRET}}
\end{center}
\end{figure}

Figure~\ref{fig:C_l_EGRET} shows the angular power spectrum of the CGB
from blazars predicted for EGRET. 
The dotted lines show the Poisson term [Eq.~(\ref{eq:pois})], 
the dashed lines show the correlation part [Eq.~(\ref{eq:corr})]
evaluated with $b_B = 1$, and the solid lines show the total.
While the blazar bias could perhaps vary from $\sim 1$ to 5, the
Poisson term dominates the angular power spectrum at all multipoles if
$b_B = 1$.
The dominance of Poisson term is due to a relatively small number of
bright blazars just below EGRET's sensitivity. The Poisson term will
decrease as we remove more fainter objects.
We also remark that the $C_l$ does not depend on the gamma-ray energy,
since we here assume all the blazars have the power-law spectrum with
the same spectral index of $\alpha_\gamma = 2.2$, and this energy
dependence exactly cancels when we divide by the mean intensity squared
$\langle I_N(E) \rangle^2$ to obtain the {\it normalized} power spectrum
$C_l$ [see Eqs.~(\ref{eq:pois}) and (\ref{eq:corr})].

\begin{figure}
\begin{center}
\includegraphics[width=8.5cm]{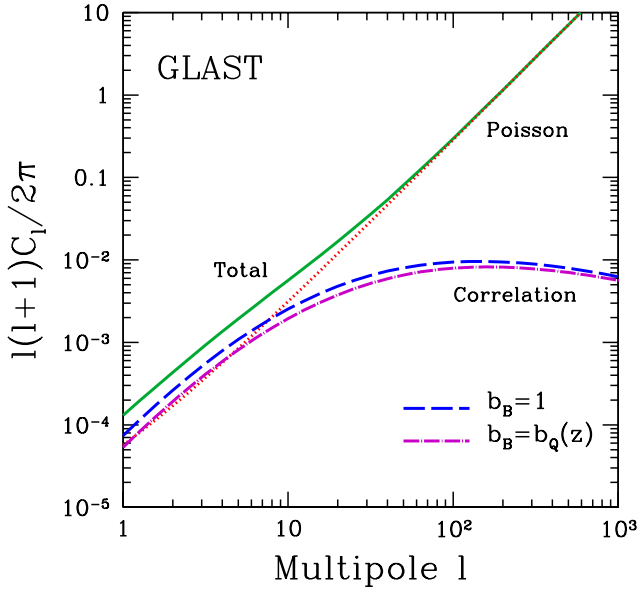}
\caption{The same as Fig.~\ref{fig:C_l_EGRET} but for the CGB anisotropy
 expected from GLAST data. \label{fig:C_l_GLAST}}
\end{center}
\end{figure}

For GLAST, we choose the point source flux limit of $2 \times
10^{-9}$ cm$^{-2}$ s$^{-1}$ ($E > 100$ MeV), $\sim 50$ times better than
EGRET, which is expected to be achieved after two years of all-sky
survey observations of sources with a spectral
index of 2~\cite{Gehrels1999}.
Our predictions for $C_l$ from GLAST data are shown in
Fig.~\ref{fig:C_l_GLAST}. 
As GLAST can detect and remove more fainter objects than EGRET, 
the Poisson term is greatly reduced while the
correlation part is almost unchanged.
If the blazar bias is larger than 1, the correlation part
would dominate the angular power spectrum at low $l$'s, which would
allow us to measure the average bias of unresolved blazars.

We also show the correlation part of the angular power spectrum
using a bias model which was inferred from the optical quasar
observations~\cite{Croom2005,Myers2006}:
\begin{equation}
 b_Q(z) = 0.53 + 0.289 (1+z)^2.
  \label{eq:QSO bias}
\end{equation}
If the unification picture of the AGNs is correct, then it may be
natural to set $b_B = b_Q(z)$.
The results from this calculation are shown as
the dot-dashed curves in Figs.~\ref{fig:C_l_EGRET} and
\ref{fig:C_l_GLAST}.
We find that these results are quite similar to the case of $b_B = 1$.
This is because at low redshift, $z \alt 0.5$, the quasar bias is close
to 1, and the main contribution to the CGB from blazars comes also from
relatively low-redshift range.
Once again, we note that the quasar bias [Eq.~(\ref{eq:QSO bias})] is
significantly different from the bias inferred from the X-ray AGN
observation, which indicated stronger
clustering~\cite{Yang2003,Basilakos2005,Gandhi2006}.
Therefore, one should keep in mind that a wide range of the blazar bias,
possibly up to $\sim 5$, is still allowed.
Hereafter, we adopt $b_B = 1$ as our canonical model, and note that
$C_l^C$ simply scales as $b_B^2$.

\section{Distinguishing dark matter annihilation and blazars}
\label{sec:Distinguishing dark matter annihilation and blazars}

The main goal in this paper is to study how to distinguish CGB
anisotropies from dark matter annihilation and from blazars.
The current uncertainty in the blazar bias would be the source 
of systematic errors, but
this can be reduced significantly by several approaches, such as the
upgraded and converged bias estimations of AGNs from the other
wavebands, direct measurement of the blazar bias from the detected point
sources by GLAST~\cite{Ando2006b}, and the CGB anisotropy at different
energies where the contribution from dark matter annihilation is likely
to be small.

\subsection{Formulation for the two-component case}
\label{sub:Formulation for the two-component case}

The total CGB intensity is the sum of dark matter annihilation and blazars:
\begin{eqnarray}
 I_{\rm CGB}(E,\bm{\hat n}) &=& I_B(E,\bm{\hat n})
  + I_{D}(E,\bm{\hat n}),
 \label{eq:intensity: B and D}\\
 \langle I_{\rm CGB}(E) \rangle &=& \langle I_B(E) \rangle
  + \langle I_D(E) \rangle,
 \label{eq:mean intensity: B and D}
\end{eqnarray}
where the subscripts $B$ and $D$ denote blazar and dark matter
components, respectively. 
The expansion coefficients of the spherical harmonics are given by
\begin{eqnarray}
 a_{lm}^{\rm CGB} & = &
  \int d\Omega_{\bm{\hat n}}\
  \frac{I_{\rm CGB}(E,\bm{\hat n}) - \langle I_{\rm CGB}(E) \rangle}
  {\langle I_{\rm CGB}(E) \rangle} Y_{lm}^\ast (\bm{\hat n})
 \nonumber\\&=&
 \int d\Omega_{\bm{\hat n}}\
 \frac{\delta I_B(E,\bm{\hat n}) + \delta I_D(E,\bm{\hat n})}
 {\langle I_{\rm CGB}(E) \rangle} Y_{lm}^\ast (\bm{\hat n})
 \nonumber\\
 & \equiv & f_B a_{lm}^{B} + f_D a_{lm}^{D},
\end{eqnarray}
where $\delta I_{B,D} \equiv I_{B,D} - \langle I_{B,D} \rangle$,
$f_{B,D} \equiv \langle I_{B,D} \rangle / \langle I_{\rm CGB} \rangle$.
These $f_B$ and $f_D$ are the fraction of contribution from the blazars
and dark matter annihilation to the total CGB flux, and we have the
relation $f_B + f_D = 1$.
Therefore, $a_{lm}^{B,D}$ is defined as the coefficient of the spherical
harmonic expansion {\it if} each component is the only constituent of
the CGB flux, the same definition as in the previous sections or of
AK06.
The total angular power spectrum, $C_l^{\rm CGB} = \langle |a_{lm}^{\rm CGB}|^2
\rangle$, is therefore written as
\begin{equation}
 C_l^{\rm CGB} = f_B^2 C_{l,B} + f_D^2 C_{l,D}
  + 2 f_B f_D C_{l,BD},
  \label{eq:two-component power spectrum}
\end{equation}
where $C_{l,B}$ and $C_{l,D}$ are the angular power spectrum of the CGB
from blazars (Sec.~\ref{sec:Cosmic gamma-ray background anisotropy II:
blazars}) and dark matter annihilation (Sec.~\ref{sec:Cosmic gamma-ray
background anisotropy I: dark matter substructure} and AK06),
respectively, and $C_{l,BD} \equiv \langle a_{lm}^B a_{lm}^{D\ast}
\rangle$ is a cross correlation term.
This cross correlation term is derived in Appendix~\ref{sec:Cross
correlation between dark matter annihilation and blazars}, and is again
divided into 1-halo and 2-halo terms, i.e.,
\begin{equation}
 C_{l,BD} = C_{l,BD}^{1h} + C_{l,BD}^{2h},
  \label{eq:cross}
\end{equation}
where each term is given by
\begin{eqnarray}
 C_{l,BD}^{1h}
  &=&
  \int dr\ \frac{W([1+z]E,z)}{\langle I_B(E)\rangle\langle
  I_D(E)\rangle}
  \int d{\cal L}\ \Phi_E({\cal L},z)
  \nonumber\\&&{}\times
  {\cal F}_E({\cal L},z)
  \frac{\langle N|M({\cal L}) \rangle}{\bar n_{\rm sh}(z)}
  u\left(\frac{l}{r},z; M[{\cal L}]\right),
  \label{eq:cross 1-halo}\\
 C_{l,BD}^{2h}
  &=&
  \int dr\
  \frac{W([1+z]E,z)}{\langle I_B(E)\rangle\langle I_D(E)\rangle}
  \int d{\cal  L}\ \Phi_E({\cal L},z)
  \nonumber\\&&{}\times  
  {\cal F}_E({\cal L},z)
  b_B({\cal L},z) \int dM\ \frac{dn(M,z)}{dM}
  \frac{\langle N|M \rangle}{\bar n_{\rm sh}(z)}
  \nonumber\\&&{}\times  
  b(M,z)
  u\left(\frac{l}{r},z;M\right)
  P_{\rm lin}\left(\frac{l}{r},z\right),
  \label{eq:cross 2-halo}
\end{eqnarray}
for the subhalo-dominated annihilation, and 
\begin{eqnarray}
 C_{l,BD}^{1h}
  &=&
  \int dr\ \frac{W([1+z]E,z)}{E\langle
  I_B(E)\rangle\langle I_D(E)\rangle}
  \int d{\cal L}\ \Phi_E({\cal L},z)
  \nonumber\\&&{}\times
  {\cal F}_E({\cal L},z)
  \frac{M[{\cal L}]}{\Omega_m\rho_c}
  v\left(\frac{l}{r},z; M[{\cal L}]\right),
  \label{eq:cross 1-halo 2}\\
 C_{l,BD}^{2h}
  &=&
  \int dr\
  \frac{W([1+z]E,z)}{E\langle I_B(E)\rangle\langle I_D(E)\rangle}
  \int d{\cal  L}\ \Phi_E({\cal L},z)
  \nonumber\\&&{}\times  
  {\cal F}_E({\cal L},z)
  b_B({\cal L},z) \int dM\ \frac{dn(M,z)}{dM} \frac{M}{\Omega_m\rho_c}
  \nonumber\\&&{}\times  
  b(M,z)
  v\left(\frac{l}{r},z;M\right)
  P_{\rm lin}\left(\frac{l}{r},z\right),
  \label{eq:cross 2-halo 2}
\end{eqnarray}
for the host-halo-dominated annihilation; $v(k,z)$ is the Fourier
transform of $\rho^2(r|M) / M\Omega_m\rho_c$.
Here we note that the function $W(E,z)$ in the latter expressions
[Eqs.~(\ref{eq:cross 1-halo 2}) and (\ref{eq:cross 2-halo 2})] is
different from Eq.~(\ref{eq:window function}), but is given by Eq.~(5)
of AK06.
In order to evaluate the 1-halo term, we need a relation between blazar
luminosity $\mathcal L$ and its host mass $M$, for which we use Eq.~(22)
of Ref.~\cite{Ando2006b}.

\subsection{Anisotropy due to dark matter annihilation in the
  two-component case}
\label{sub:Anisotropy due to dark matter annihilation}

Since our main thrust here is how to detect the dark matter annihilation
signature out of the CGB in the presence of more common (and plausibly known) blazar component,
we focus our attention only on the energy of 10 GeV, a typical energy of
gamma rays expected from the annihilation of dark matter particles with
a mass of $\sim 100$ GeV.
From now on we treat the blazar contribution as the ``background noise.''
A signal and background of the angular power spectrum is, respectively,
\begin{eqnarray}
 C_l^{s} &=& f_D^2 C_{l,D} + 2 f_D (1-f_D) C_{l,BD},
  \label{eq:signal}\\
 C_l^{b} &=& (1-f_D)^2 C_{l,B},
  \label{eq:background}
\end{eqnarray}
and we assume that the background is very well known.
This is a reasonable assumption, provided that
we can model the GLF and bias of unresolved sources from
those of resolved (detected) sources in the GLAST data.
One can also calibrate the background (blazar) component using the angular power
spectrum of the CGB at lower energies such as 100 MeV, where
contribution from the dark matter (neutralinos) annihilation may be ignored.

We use several LDDE parameter sets for the blazar GLF, which are listed
in Table~\ref{table:LDDE model}.
The faint end of the GLF, which makes the largest contribution to the
CGB, is given by $\rho_\gamma\propto \kappa L^{-\gamma_1}_\gamma$,
where $\gamma_1$ is the faint-end slope and $\kappa$ is the overall
normalization~\cite{Narumoto2006}. 
We vary these parameters such that the blazars explain a certain
fraction, $f_B^{\rm EGRET}$,  of the CGB intensity measured by the EGRET
at 10~GeV.
We call these models LDDE10, LDDE30, LDDE50, and LDDE70 for
$f_B^{\rm EGRET}=0.1$, 0.3, 0.5, and 0.7, respectively.
We note that the best-fitting LDDE model, $(\gamma_1,\kappa) = (1.19,
5.11\times 10^{-6})$, explains only $\sim 20$\% of the CGB flux at 10
GeV, and therefore, one should keep in mind that this list includes
models that are somewhat disfavored from the redshift and luminosity
distributions of the EGRET blazars.
In the two-component treatment, the dark matter fraction is obviously
obtained by $f_D = 1 - f_B$.

As for the GLAST data, 
the
contribution to the CGB from blazars will be greatly reduced as 
GLAST can resolve and remove more blazars.
On the other hand, the contribution from dark matter annihilation will
not change because it is unlikely that we can detect individual dark
matter halos via annihilation even with
GLAST~\cite{Oda2005,Horiuchi2006}. 
Therefore, the total CGB intensity, $\langle I_{\rm CGB} \rangle$,
that would be measured by GLAST would be smaller than that measured
by EGRET, while keeping intensity from dark matter
annihilation unchanged.
The expected fractional contributions to the CGB in the GLAST data from blazars 
or dark matter annihilation are
also shown in Table~\ref{table:LDDE model}.

\begin{table}
\begin{center}
\caption{Blazar LDDE GLF models adopted in the two-component (dark
 matter annihilation and blazars) study of
 the CGB anisotropy. Values of $\gamma_1$ (the faint-end slope of GLF)
 and $\kappa$ (the overall normalization of GLF) specify each 
 model, and the output is given in terms of the fractional contributions
 to the CGB from  blazars, $f_B$, and dark matter annihilation, $f_D$,
 for EGRET or GLAST. \label{table:LDDE model}}
\begin{tabular}{lccccc}
 \hline\hline
 Model & $(\gamma_1,10^6\kappa)$ & $f_B^{\rm EGRET}$ & $f_D^{\rm EGRET}$
 &  $f_B^{\rm GLAST}$ & $f_D^{\rm GLAST}$\\
 \hline
 LDDE10 & $(1.05,6.33)$ & 0.1 & 0.9 & 0.03 & 0.97\\
 LDDE30 & $(1.23,4.72)$ & 0.3 & 0.7 & 0.20 & 0.80\\
 LDDE50 & $(1.29,4.11)$ & 0.5 & 0.5 & 0.39 & 0.61\\
 LDDE70 & $(1.33,3.69)$ & 0.7 & 0.3 & 0.61 & 0.39\\
 \hline\hline
\end{tabular}
\end{center}
\end{table}

\begin{figure*}
\begin{center}
\rotatebox{-90}{\includegraphics[width=12cm]{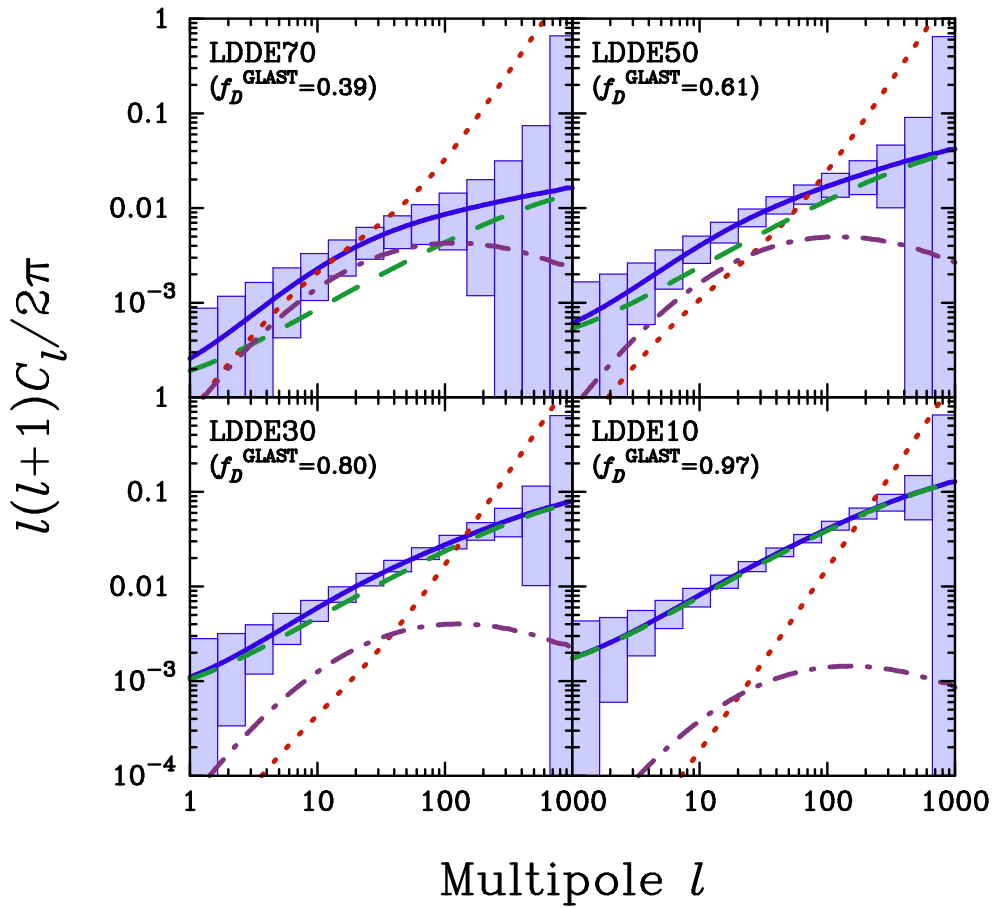}}
\caption{Angular power spectrum of the CGB from dark matter annihilation
 ($f_D^2 C_{l,D}$; dashed), blazars ($f_B^2 C_{l,B}$; dotted), and cross
 correlation ($2f_B f_D C_{l,BD}$; dot-dashed) that would be measured by
 GLAST  at $E = 10$ GeV, for various models of the blazar GLF and various
 fractions of dark matter contribution to the CGB, $f_D$
 (Table~\ref{table:LDDE model}). The adopted dark matter mass is 100
 GeV and the gamma-ray emission is assumed to be dominated by the
 substructure. The total signal, $C_l^s=f_D^2 C_{l,D}+2f_B f_D
 C_{l,BD}$, is 
 shown as the solid curve, while the corresponding GLAST errors ($\delta
 C_l^s$; for two
 years) are indicated as boxes. The signal is to be detected if it is
 larger than the size of
 errors ($C_l^s > \delta C_l^s$). The subhalo distribution in a halo of
 mass $M$ is assumed to be $\langle N|M \rangle \propto
 M$.  \label{fig:C_l_subhalo}}
\end{center}
\end{figure*}

\begin{figure*}
\begin{center}
\rotatebox{-90}{\includegraphics[width=12cm]{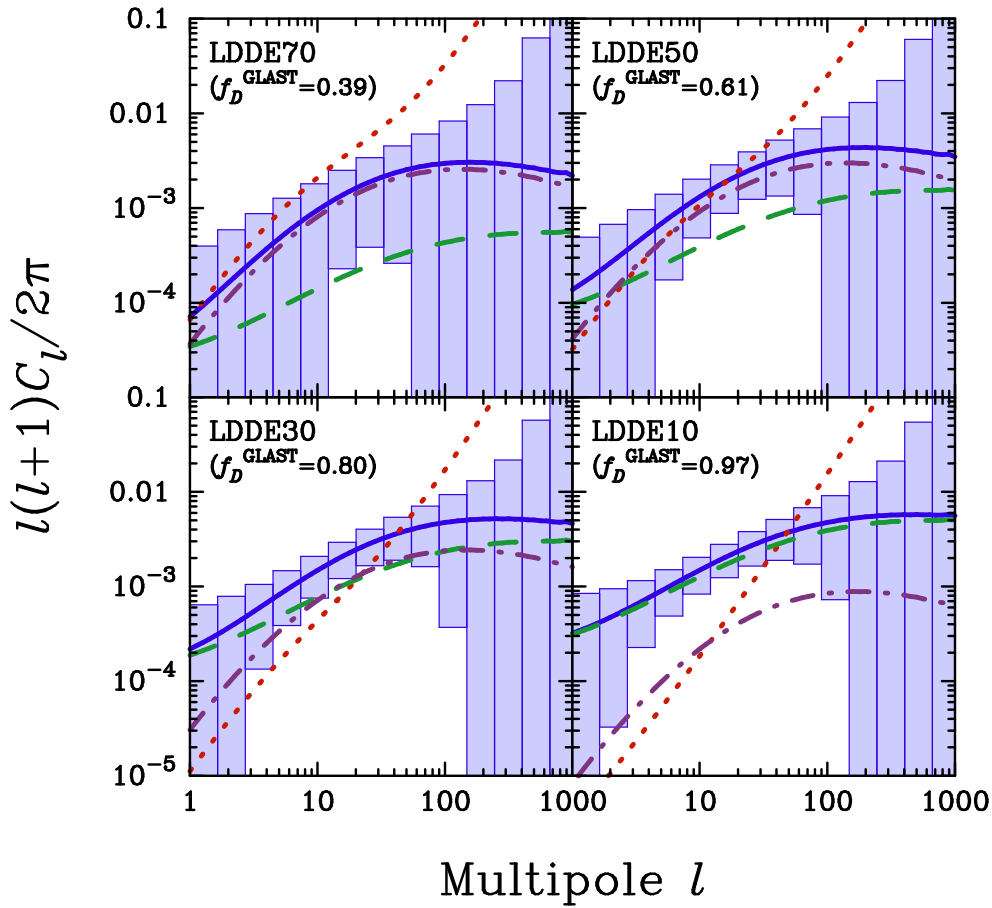}}
\caption{The same as Fig.~\ref{fig:C_l_subhalo} but for $\langle N|M
 \rangle \propto M^{0.7}$. \label{fig:C_l_subhalo_index7}}
\end{center}
\end{figure*}

In Figs.~\ref{fig:C_l_subhalo} and \ref{fig:C_l_subhalo_index7}, we show
the angular power spectrum of dark matter annihilation
(i.e., $C_l^s$; solid) as well as of background ($C_l^b$; dotted) that
would be measured by GLAST, for the blazar models shown in
Table~\ref{table:LDDE model}. We assume that the dark matter signal is
dominated by substructures. We also show the expected 1-$\sigma$
error bands of $C_l^s$. Note that the error bands do include contributions
from the blazar power. (See Sec.~\ref{sub:Detectability of the dark
matter component} for the error estimation.)
The subhalo distribution in each halo is assumed to be $\langle N|M
\rangle \propto M$ (Fig.~\ref{fig:C_l_subhalo}) and 
$\propto M^{0.7}$ (Fig.~\ref{fig:C_l_subhalo_index7}).
The shape
of the angular power spectrum is quite different between dark matter
annihilation and blazars, a characteristic that should be useful for a
smoking gun detection of dark matter annihilation.

Even though the dark matter contribution to the CGB 
is relatively small for $\alpha = 0.7$, the
cross correlation term (due mainly to 2-halo term) is still reasonably
large; we find that the 1-halo term of the cross correlation is
always negligibly small as long as an empirical luminosity--host mass
relation is used (Eq.~(22) of Ref.~\cite{Ando2006b}).
Although the shape of the cross correlation is similar to the pure
blazar correlation term (because both are proportional to the linear
power spectrum $P_{\rm lin}$), its normalization gives us useful clue of
the source because the normalization of the pure blazar term should be
known to reasonable accuracy in advance.

\begin{figure*}
\begin{center}
\rotatebox{-90}{\includegraphics[width=12cm]{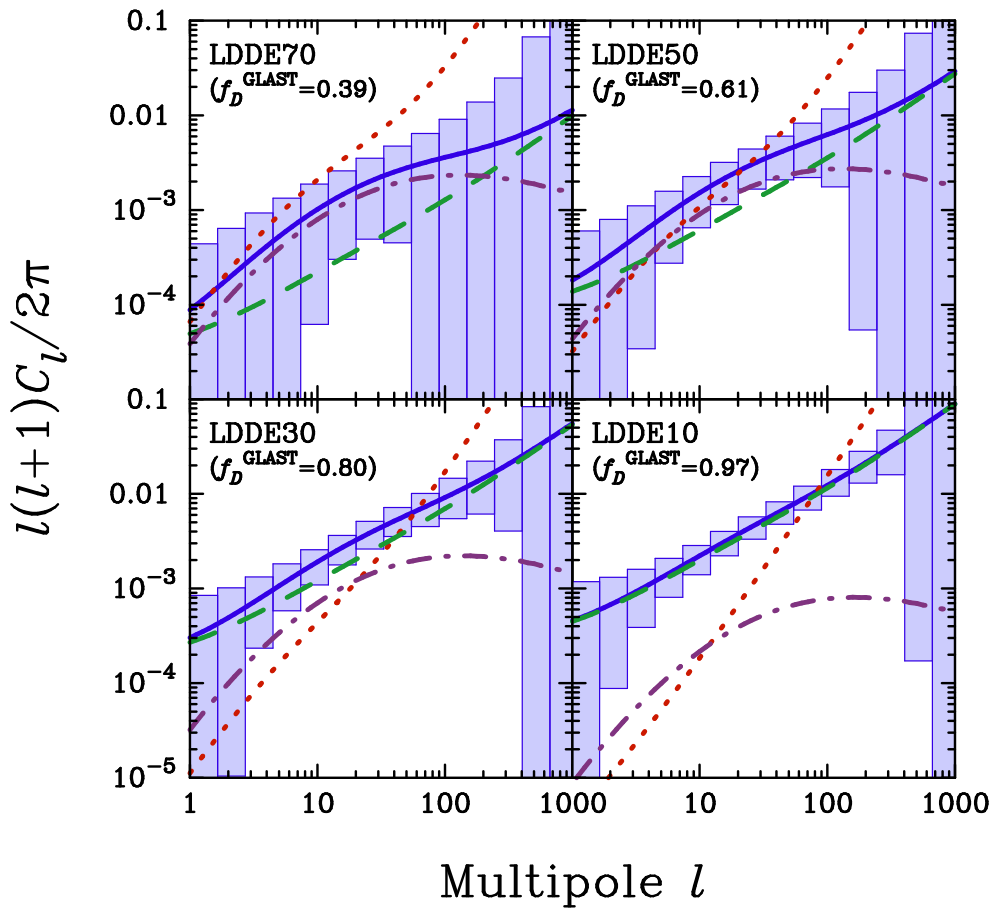}}
\caption{The same as Fig.~\ref{fig:C_l_subhalo}, but in the case of the
 host-halo-dominated annihilation (no substructures; AK06). The minimum
 mass of dark 
 matter halos is assumed to be $10^{6}
 M_\odot$. \label{fig:C_l_halo_cut6}}
\end{center}
\end{figure*}

\begin{figure*}
\begin{center}
\rotatebox{-90}{\includegraphics[width=12cm]{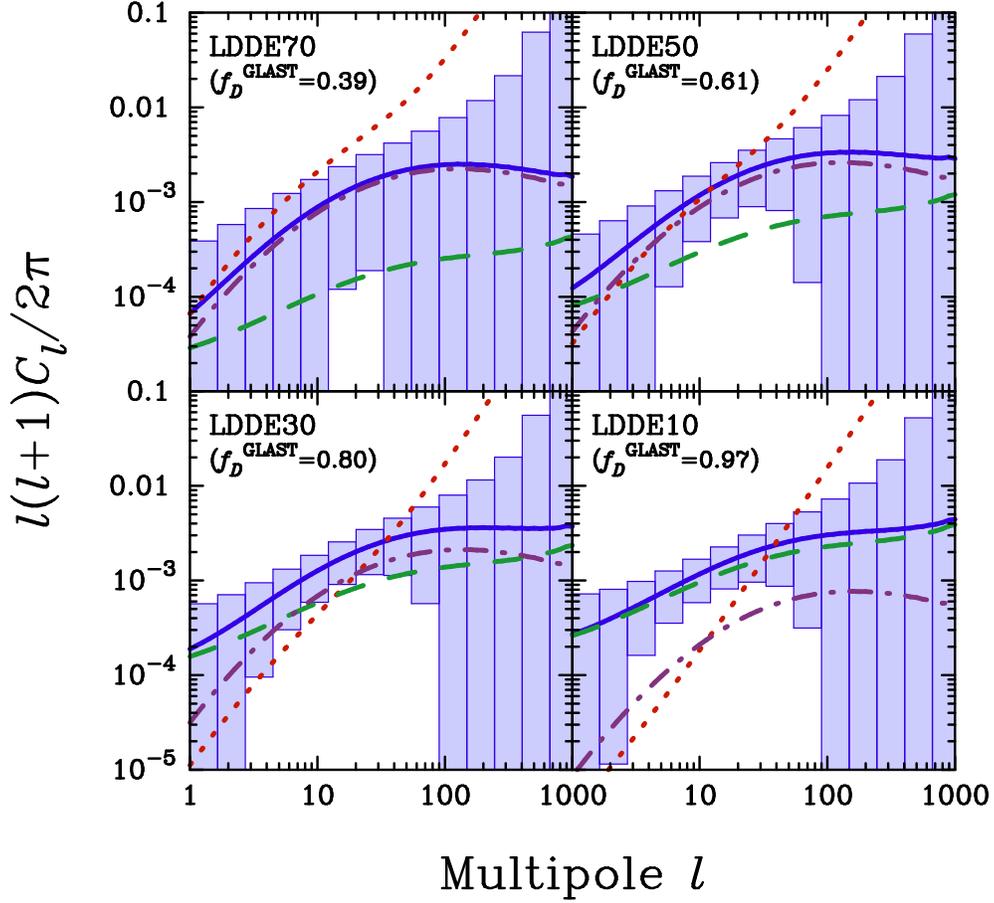}}
\caption{The same as Fig.~\ref{fig:C_l_halo_cut6} but with minimum halo
 mass of $10^{-6} M_\odot$. \label{fig:C_l_halo}}
\end{center}
\end{figure*}

We then repeat the same analysis for the host-halo-dominated
annihilation (no substructures).
This is a straightforward extension of the study in AK06, to which the
background is added.
We show in Figs.~\ref{fig:C_l_halo_cut6} and \ref{fig:C_l_halo} the 
power spectra for $M_{\rm min} = 10^6 M_\odot$ and $10^{-6}
M_\odot$, respectively.
The smaller minimum mass is motivated by the
free-streaming scale of neutralinos, $10^{-6} M_\odot$
\cite{Green2005,Loeb2005,Profumo2006}, while the larger one 
represents an extreme case in which halos smaller than a million solar
masses ($M < 10^6 M_\odot$) have been tidally disrupted.
(Note, however, that there still remains large allowed range for the
former case, $10^{-12}$--$0.1 M_\odot$~\cite{Profumo2006}).
Compared with the subhalo-dominated case, the anisotropy signature is
typically smaller, but the general tendency is almost the same,
justifying qualitative arguments regarding substructures given in AK06. 

\subsection{Can GLAST detect dark matter annihilation?}
\label{sub:Detectability of the dark matter component}

We use the standard procedure to calculate the projected 1-$\sigma$
error (binned over $\Delta l$) 
on the extracted power spectrum of the CGB from dark matter
annihilation: 
\begin{equation}
 \delta C_l^s =
  \sqrt{\frac{2}{(2l+1)\Delta l f_{\rm sky}}}
  \left(C_l^s+C_l^b+\frac{C_N}{W_l^2}\right),
  \label{eq:C_l_error}
\end{equation}
where $f_{\rm sky} = \Omega_{\rm sky} / 4 \pi$ is the fraction of the
sky covered by GLAST, $\Delta l$ is the bin width (which we shall take
to be $\Delta l = 0.5l$), $C_N = \Omega_{\rm
sky} N_{\rm total} / N_{\rm CGB}^2$ is the power spectrum of photon
noise, and $N_{\rm CGB}$ and $N_{\rm total}$ are the photon numbers of the
CGB and total (CGB plus other backgrounds), respectively, 
expected from the region
$\Omega_{\rm sky}$, and $W_l$ is the window function of a Gaussian point
spread function, $W_l = \exp (-l^2 \sigma_b^2 / 2)$.
Note that this formula assumes that CGB anisotropy obeys Gaussian
statistics. 
We take the following specifications for GLAST: the field of view is
$\Omega_{\rm fov} = 2.4$ sr, the angular resolution is $\sigma_b =
0.115^\circ$, and the effective area is $A_{\rm eff} = 10^4$ cm$^2$,
both evaluated at $E = 10$ GeV~\cite{Gehrels1999}.
We assume a two-year all-sky survey ($T = 2$ yr), which corresponds to
mean exposure of $A_{\rm eff}t_{\rm eff} = A_{\rm eff} T \Omega_{\rm
fov} / 4\pi = 1.2 \times 10^{11}$ cm$^2$ s, towards each point in the
sky.

Equation~(\ref{eq:C_l_error}) clearly shows that the ``astrophysical
background noise'' 
from blazars, $C_l^b$, contributes to the error budget. 
On the other hand, the background that contributes to $C_N$ includes
detector noise and Galactic gamma-ray radiation, which we call foreground.
The detector noise is very small for GLAST, about 5\% of the CGB flux
above 100 MeV; this decreases significantly as the gamma-ray energy and
the detector noise is safely assumed to be negligible.
The Galactic foreground is more difficult to estimate, but according
to recent calculations using a numerical code of the Galactic cosmic-ray
propagation, its intensity at 10 GeV is estimated to be $E^2 I_{N,{\rm
gal}} \simeq 10^{-7}\ \mathrm{GeV\ cm^{-2}\ s^{-1}\ sr^{-1}}$ at high
Galactic latitudes, $|b| > 20^\circ$, which is an order of magnitude
smaller than the observed CGB intensity (see Fig.~4 of
Ref.~\cite{Strong2004b}).
Below this latitude, the Galactic foreground entirely dominates the
gamma-ray flux; thus, we do not consider $|b| < 20^\circ$ further.
We also note that the Galactic foreground may contribute to the $C_l$
due to spatial fluctuation of the sources, but we neglect this effect in
the following arguments. 
We postulate that this is a reasonable approximation because its flux at
the large Galactic latitude is very small compared with the CGB, which
suppresses the contribution to the total $C_l$ as well; 
including this effect requires a detailed modeling of the Galactic
cosmic-ray propagation, which is beyond the scope of the present
study.
The fraction of the sky relevant to our analysis is therefore $f_{\rm sky} =
1-\cos 70^\circ = 0.66$, which is reasonably close to 1. We do not
lose much by this Galactic cut.
The number of CGB photons expected from this region and for $T = 2$ yr
is $N_{\rm CGB} \simeq E I_{\rm CGB} A_{\rm eff} t_{\rm eff} \Omega_{\rm
sky} = 10^5 (E/10\ {\rm GeV})^{-1}$, while the corresponding photon
count from foreground is $\sim 10^4$---negligible compared with
that of the CGB.
We finally obtain $C_N \simeq \Omega_{\rm sky}/ N_{\rm CGB} = 8\times 10^{-5}
(E/10\ {\rm GeV})$ sr.
Note that this number is based upon the CGB intensity measured from
EGRET, and it will be smaller for GLAST because GLAST can detect and
remove more fainter blazars. 
While this reduction is eventually found to give no significant effect,
we take more precise approach in the actual error estimation. Here,
just in order to provide a rough idea of the size of each quantity, we
used the CGB flux obtained with EGRET.

%
%

We show our predictions with the error bars calculated from
Eq.~(\ref{eq:C_l_error}) in Figs.~\ref{fig:C_l_subhalo}--\ref{fig:C_l_halo}.
We find that,
for {\it any models} we have considered here, one should be able to
detect the angular power
spectrum from dark matter annihilation with GLAST
in two years of operation, as long as the dark matter contribution to the
mean CGB flux is greater than 30\% at some energy within the GLAST
energy window. 
This statement is independent of the density profile adopted as it is
derived with the {\it guaranteed} power spectrum, the 2-halo term.

\section{Discussion and conclusions}
\label{sec:Discussion and conclusions}

In this paper we have presented detailed calculations of  the angular
power spectrum of CGB 
anisotropy from dark matter annihilation in cosmological dark matter
halos as well as from {\it unresolved} blazars.
The power spectrum of dark matter annihilation from smooth NFW halos
(i.e., no substructures) has been calculated by AK06~\cite{Ando2006a}, and
the power spectrum of {\it resolved} (detected) blazars has been
calculated by Ref.~\cite{Ando2006b}.
Our work builds on and extends these results by taking into account the
effects of dark matter substructures explicitly, by means of the Halo
Occupation Distribution of subhalos.
These calculations should provide a useful benchmark for the
angular power spectrum of CGB that would be measured by GLAST.

Our results are very encouraging: 
one should be able to detect the angular power
spectrum from dark matter annihilation with GLAST, whether dark matter
halos are smooth or clumpy,  as long as the dark matter contribution to the
mean CGB flux is greater than 30\% at some energy within the GLAST
energy window. 
This is a rather modest requirement given the fact that
unresolved blazars appear to contribute to the mean CGB only at the 25--50\%
level \cite{Narumoto2006}.
As far as the mean CGB is concerned it has been pointed out that 
subhalos are necessary in order for dark
matter annihilation to make a  significant contribution without violating
stringent constraints from the Galactic Center~\cite{Ando2005,Oda2005,Horiuchi2006}. Thus, 
our current ``best'' predictions are either Fig.~\ref{fig:C_l_subhalo}
or Fig.~\ref{fig:C_l_subhalo_index7}, depending on the degree to which
tidal disruption of subhalos is important.

The ``background noise'' for dark matter annihilation searches
includes the blazar anisotropy, detector noise, and the Galactic foreground.
The blazar anisotropy should be well calibrated with the CGB anisotropy
at lower energies (where dark matter signal is probably unimportant), analysis of AGNs selected with other wavebands, and/or
the power spectrum of resolved point sources.
The detector noise is always negligible.
The Galactic foreground contamination is serious near the plane,
$|b|<20^\circ$, whereas its flux is found to be at most 
$\sim 10$\% of the CGB flux at high Galactic latitudes.
The anisotropy due to the foreground is thus accordingly reduced.
We also note that dark matter annihilation in the Galactic halo may also
give a contribution to the CGB flux comparable to that from the
cosmological halos (e.g., Ref.~\cite{Oda2005,Diemand2006}).
Adding this component would increase the predicted
anisotropy from dark matter annihilation.

GLAST covers energy spectrum up to 300 GeV, which might enable us to
probe a line signature in the CGB due to direct annihilation of dark
matter particles into two photons (or one photon plus one $Z^0$ boson).
In addition to quite robust spectral feature, one could also use the
CGB anisotropy as a consistency check.
When the spectrum has a feature as the case of lines, the angular power
spectrum becomes larger~\cite{Zhang2004,Ando2005}, which makes this
method even more promising.

Finally, we comment on how our conclusions might be affected
by the other astrophysical sources (of either known or unknown species)
that we have not considered. 
If the gamma-ray emitter is a point source and follows the distribution
of dark matter halos, then the number of such sources contributing to
the CGB is the only parameter that determines 
the amplitude of the power spectrum. 
Whatever the sources are, they should be dimmer than blazars on average;
we must have seen them in the EGRET data otherwise.
If the number of such sources is much larger than blazars, 
then the Poisson
term, $C_l^P$, is highly suppressed and is unimportant.
If these sources are rare, forming only in large halos, the power
spectrum (both 1-halo and 2-halo terms) can be large, but it is
extremely difficult for them to give a dominant contribution to the
CGB.


If these sources are spatially extended, e.g., clusters of galaxies,
then the shape of the angular power spectrum depends on their gamma-ray 
profile.
How the galaxy clusters are extended in gamma rays depends on
the emission mechanism, which is uncertain. (No direct detection of
gamma rays towards known clusters has been made so far.)
Incidentally one can make the most conservative (or pessimistic)
prediction as to how much the unresolved clusters would contribute to
CGB anisotropy, by assuming that the galaxy clusters are point sources.
We have performed the same calculations for galaxy clusters
for both the proton-proton
collision~\cite{Colafrancesco1998} and inverse-Compton
models~\cite{Totani2000}.
Our results again suggest that the dark matter component should be detectable
significantly, as long as its contribution to the EGRET CGB is more than
$\sim 30$\%.
Angular distribution of background radiation from galaxy clusters has
been investigated also in previous papers, for radio
waveband~\cite{Waxman2000,Keshet2004} as well as for gamma
rays~\cite{Waxman2000,Keshet2003}.


Based upon these results, we conclude that anisotropy in the CGB that
would be measured by GLAST has to be analyzed in search for signatures of
dark matter annihilation. If detected, it should provide us with the
first (albeit indirect) evidence for emission from dark matter
particles, which would shed light on the nature of the still-mysterious 
dark matter in the universe.

\acknowledgments

S.A. and E.K. would like to thank Jennifer Carson for useful
discussions.
S.A. is also grateful to Edward Baltz, Marc Kamionkowski, Stefano
Profumo, and Lawrence Wai for comments.
S.A. was supported by Sherman Fairchild Foundation at Caltech.
E.K. acknowledges support from the Alfred P. Sloan Foundation.
T.N. and T.T. were supported by a Grant-in-Aid for the 21st Century
COE ``Center for Diversity and Universality in Physics'' from the
Ministry of Education, Culture, Sports, Science and Technology of
Japan.

\appendix


\section{Relation between angular and spatial power spectrum}
\label{ap:Relation between angular and spatial power spectrum}

\subsection{Dark matter substructure}
\label{apsub:Dark matter substructure}

In this subsection, we derive Eq.~(\ref{eq:Limber's equation}) following
the halo model approach~\cite{Scherrer1991,Seljak2000}.
Since the gamma-ray intensity due to annihilation dominated by subhalos
depends on the density, which can be written as a superposition of
densities in different halos, we obtain the following expression for the
volume emissivity:
\begin{eqnarray}
 P_\gamma (E,\bm x) &=& \int dM_1 d^3 x_1\
  \sum_i \delta_D (M_1 - M_i) \delta_D^3 (\bm x_1 - \bm x_i)
  \nonumber\\&&{}\times
  \langle N|M \rangle u(\bm x - \bm x_1|M_1) E \mathcal N_{\rm sh}(E)
  \label{eq:volume emissivity: superposition}
\end{eqnarray}
where $\delta_D^N$ is the $N$-dimensional delta function, and $\bm x$
represents comoving coordinate.
As in Ref.~\cite{Scherrer1991}, the ensemble average of the sum over
delta functions is simply the seed density:
\begin{equation}
 \left\langle \sum_i\delta_D^3 (\bm x_1 - \bm x_i)\delta_D (M_1 -
  M_i)\right\rangle
 = \frac{dn(M_1)}{dM},
\end{equation}
and using this into Eqs.~(\ref{eq:volume emissivity: superposition}) and
(\ref{eq:intensity general}), we recovers Eq.~(\ref{eq:mean intensity:
simple form}) since $\int d^3x_1 u(\bm x- \bm x_1|M_1) = 1$.

We then define the two-point correlation function of subhalos as
\begin{equation}
 \xi_{\rm sh}(\bm x - \bm y) =
  \frac{\langle \delta P_\gamma (\bm x) \delta P_\gamma (\bm y) \rangle}
  {\langle P_\gamma (|\bm x|)\rangle \langle P_\gamma (|\bm y|)\rangle},
  \label{apeq:subhalo correlation}
\end{equation}
where $\delta P_\gamma (\bm x) = P_\gamma (\bm x) - \delta P_\gamma
(|\bm x|)$.
Now we evaluate $\langle P_\gamma (\bm x) P_\gamma (\bm y)\rangle$.
The ensemble average of the product of seed densities is written as
follows~\cite{Scherrer1991}:
\begin{widetext}
\begin{eqnarray}
 \lefteqn{\left\langle \sum_i\delta_D^3 (\bm x_1 - \bm x_i)\delta_D (M_1
 - M_i) \sum_j\delta_D^3 (\bm x_2 - \bm x_j)\delta_D (M_2 - M_j)
 \right\rangle}\nonumber\\
 &=& \frac{dn(M_1)}{dM} \frac{dn(M_2)}{dM}
  \left[ 1 + \xi_s^{(2)}(\bm x_1 - \bm x_2; M_1, M_2)\right]
  + \frac{dn(M_1)}{dM} \delta_D^3 (\bm x_2 - \bm x_1)
  \delta_D (M_2 - M_1),
\end{eqnarray}
\end{widetext}
where $\xi_s^{(N)}$ is the $N$-point correlation function of the seed.
The first term represents the two-halo contribution, i.e., two points
considered, $\bm x_1$ and $\bm x_2$, are in two distinct halos, while
the second are the one-halo contribution where these two points are in
the same halo.
By substituting this expression into Eq.~(\ref{eq:volume emissivity:
superposition}), and using the relation $\langle \delta P_\gamma (\bm x)
\delta P_\gamma (\bm y) \rangle = \langle P_\gamma (\bm x) P_\gamma (\bm
y) \rangle - \langle P_\gamma (|\bm x|)\rangle \langle P_\gamma (|\bm
y|)\rangle$ and Eq.~(\ref{eq:mean volume emissivity}), we obtain
\begin{eqnarray}
 \xi_{\rm sh}(\bm x - \bm y)
  &=& \xi_{\rm sh}^{1h}(\bm x - \bm y)
  + \xi_{\rm sh}^{2h}(\bm x - \bm y)
  \nonumber\\
  &=& \int d^3x_1 dM_1\ \frac{dn(M_1)}{dM}
   \left(\frac{\langle N|M \rangle}{\bar n_{\rm sh}}\right)^2
   \nonumber\\&&{}\times
   u(\bm x - \bm x_1 | M_1) u(\bm y - \bm x_1 | M_1)
   \nonumber\\&&+
  \int d^3x_1 d^3x_2 dM_1 dM_2\
  \frac{dn(M_1)}{dM} \frac{dn(M_2)}{dM}
  \nonumber\\&&{}\times
  \frac{\langle N|M_1 \rangle \langle N|M_2 \rangle}{\bar n_{\rm sh}^2}
  u(\bm x - \bm x_1 | M_1)
  \nonumber\\&&{}\times
  u(\bm y - \bm x_2 | M_2)
  \xi_s^{(2)}(\bm x_1 - \bm x_2; M_1,M_2).
  \nonumber\\
\end{eqnarray}
We define the power spectrum of subhalos $P_{\rm sh}(k)$ as Fourier
transforms of $\xi_{\rm sh}(r)$.
Remembering that convolution in the real space corresponds to a simple
product in the Fourier space, we obtain for the expression for $P_{\rm
sh} = P_{\rm sh}^{1h} + P_{\rm sh}^{2h}$ as follows:
\begin{eqnarray}
 P_{\rm sh}^{1h}(k) &=& \int
  dM\ \frac{dn(M)}{dM}
  \left(\frac{\langle N|M \rangle}{\bar n_{\rm sh}}\right)^2
  |u(k|M)|^2,\nonumber\\
 \\
 P_{\rm sh}^{2h}(k) &=& \left[\int
  dM\ \frac{dn(M)}{dM}
  \frac{\langle N|M \rangle}{\bar n_{\rm sh}}
  b(M) |u(k|M)|\right]
 \nonumber\\&&{}\times
 P_{\rm lin}(k),
\end{eqnarray}
where we used an approximation as $\xi_s^{(2)}(r;M_1,M_2) \approx b(M_1)
b(M_2) \xi_{\rm lin} (r)$ by introducing the bias factor $b(M)$ and
linear matter correlation function $\xi_{\rm lin}$.
These expressions are identical to Eqs.~(\ref{eq:1-halo term}) and
(\ref{eq:2-halo term}).

Finally we derive the relation between angular power spectrum $C_l$ and
spatial power spectrum $P_{\rm sh}(k)$ in the following.
We start with the angular correlation function that is defined by
\begin{equation}
 \langle I_N(E) \rangle^2 C(\theta)
  = \langle \delta I_N(\bm{\hat n}_1,E) \delta I_N(\bm{\hat
  n}_2,E)\rangle,
  \label{apeq:angular correlation function}
\end{equation}
where $\cos\theta = \bm{\hat n}_1 \cdot \bm{\hat n}_2$.
Using this definition together with Eqs.~(\ref{eq:intensity general}),
(\ref{eq:mean volume emissivity}), (\ref{eq:window function}), and
(\ref{apeq:subhalo correlation}), we get
\begin{equation}
 \langle I_N\rangle^2 C(\theta)
  = \int dr_1 dr_2\ W(z_1) W(z_2) \xi_{\rm sh}(\bm r_1 - \bm
  r_2|z_1,z_2),
  \label{apeq:angular correlation 2}
\end{equation}
where $z_1$ and $z_2$ are redshifts corresponding to $r_1$ and $r_2$,
respectively, and we suppressed energy indices for simplicity.
To further simplify, we use small separation approximation following
Ref.~\cite{Peebles1980} [we use $dr_1dr_2 = dr d\eta$, with definitions
of $r = (r_1 + r_2)/2, \eta = r_2 - r_1$], and
arrive at
\begin{equation}
 \langle I_N \rangle^2 C(\theta)
  = \int dr d\eta\ W(z)^2
  \xi_{\rm sh}(\eta \bm{\hat r} + r\bm \theta | z).
  \label{apeq:angular correlation 3}
\end{equation}
The angular power spectrum is related to the correlation function
through
\begin{equation}
 C_l = \int d^2\theta\ e^{-i\bm l\cdot \bm\theta}C(\theta),
  \label{apeq:angular power spectrum}
\end{equation}
for small scales, or for large multipoles.
Then, the Fourier transformation of the relevant quantity simplifies as
follows:
\begin{eqnarray}
 \lefteqn{
  \int d^2\theta\ e^{-i\bm l\cdot\bm\theta} \int d\eta\
  \xi_{\rm sh} (\eta\bm{\hat r} + r\theta\bm{\hat\theta},z)
  }\nonumber\\
  &=&
  \int d^2\theta\ e^{-i\bm l\cdot\bm\theta} \int d\eta
  \int\frac{d^3k}{(2\pi)^3}\ P_{\rm sh}(k,z)e^{i\bm k\cdot
  (\eta\bm{\hat r} + r\theta\bm{\hat\theta})}
  \nonumber\\&=&
  \int d^2\theta \int d\eta
  \int\frac{dk_{\parallel}d^2k_{\perp}}{(2\pi)^3}\ 
  P_{\rm sh}(k,z) e^{ik_{\parallel} \eta} e^{i\bm{\theta}\cdot
  (r\bm{k}_{\perp}-\bm l)}
  \nonumber\\&=&
  \int dk_{\parallel}d^2k_{\perp}\
  P_{\rm sh}\left(\sqrt{k_{\parallel}^2 + k_{\perp}^2},z\right)
  \delta_D (k_{\parallel})\delta_D^2 (r\bm k_{\perp} - \bm l)
  \nonumber\\&=&
  \frac{1}{r^2}P_{\rm sh}\left(k=\frac{l}{r},z\right).
  \label{apeq:fourier}
\end{eqnarray}
In the second equality, we decomposed the wavenumber $\bm k$ by the
components parallel and perpendicular to $\bm r$, i.e., $\bm k = \bm
k_{\parallel} + \bm k_{\perp}$, and used $d^3k = dk_{\parallel} d^2
k_{\perp}$.
Therefore, together with Eqs.~(\ref{apeq:angular correlation 3}) and
(\ref{apeq:angular power spectrum}), we arrive at our demanded relation,
Eq.~(\ref{eq:Limber's equation}).

\subsection{Blazars}
\label{apsub:Blazars}

In this subsection, we derive Eqs.~(\ref{eq:pois}) and (\ref{eq:corr})
following discussions in Ref.~\cite{Peebles1980} (see also,
Ref.~\cite{Komatsu1999}).
The Poisson noise $C_l^P$ is obtained by setting $\theta = 0$ for
$C(\theta)$ [Eq.~(\ref{apeq:angular correlation function})], and it is
given by Eq.~(58.14) of Ref.~\cite{Peebles1980} [note with Eq.~(58.12)
that the definition of the angular power spectrum is slightly
different], which is
\begin{eqnarray}
 E^2\langle I_N(E)\rangle^2 C_l^{P} &=&
  \int dz\ \frac{d^2V}{dzd\Omega}
  \int d{\cal L}\ \Phi({\cal L},z) \mathcal F_E({\cal L},z)^2,
  \nonumber\\
  \label{eq:pois2}
\end{eqnarray}
in our notation.
It is equivalent to Eq.~(\ref{eq:pois}).

The correlation term of the angular power spectrum $C_l^{C}$, on the
other hand, is obtained with the Fourier transformation of the angular
correlation function for $\theta \neq 0$ (Eq.~(58.6) of
Ref.~\cite{Peebles1980})
\begin{eqnarray}
 \lefteqn{
  E^2 \langle I_N(E)\rangle^2 C(\theta)
  } \nonumber\\
  &=& 
  \frac{1}{16\pi^2}\int dz\ \frac{d^2V}{dzd\Omega} \frac{1}{(1+z)^2r(z)^2}
  \int_{-\infty}^{\infty}du\
  \nonumber\\&&{}\times
  \xi_B(u\bm{\hat r} + r(z)\theta\bm{\hat\theta},z)
  \left[\int d{\cal L}\ {\cal L}\Phi_E({\cal L},z)
  \right]^2,\nonumber\\
\end{eqnarray}
where $\xi_B(r,z)$ is the two-point correlation function of blazars;
here we suppressed the luminosity index $\mathcal L$ but note that this
dependence should appear in general.
Then, using the Fourier transformation that is similar to
Eq.~(\ref{apeq:fourier}), we obtain the correlation part of the angular
power as
\begin{eqnarray}
 \lefteqn{
  E^2 \langle I_N(E) \rangle^2 C_l^{C}
  }\nonumber\\
  &=&
  \int_{\theta \neq 0} d^2\theta\ e^{-i\bm l\cdot \bm\theta}
  E^2 \langle I_N(E) \rangle^2 C(\theta)
  \nonumber\\&=&
  \int dz\ \frac{d^2V}{dzd\Omega} \left[\frac{1}{4\pi (1+z) r(z)^2}\right]^2
  P_B\left(\frac{l}{r(z)},z\right)
  \nonumber\\&&\times
  \left[\int d{\cal L}\ {\cal L} \Phi_E({\cal L},z)\right]^2
  \nonumber\\&=&
  \int dz\ \frac{d^2V}{dzd\Omega} P_B\left(\frac{l}{r(z)},z\right)
  \nonumber\\&&\times
  \left[\int d{\cal L}\ \Phi_E({\cal L},z)
  \mathcal F_E({\cal L},z)\right]^2,\nonumber\\
  \label{eq:corr2}
\end{eqnarray}
where in the last equality we used the relation $r(z) = (1+z)^{-1}
d_L(z)$ and Eq.~(\ref{eq:point source flux}).
This is very similar to Eq.~(\ref{eq:corr}).
It is obvious that if we introduced the bias parameter from the
beginning, as $\xi_B(r;{\cal L}_1,{\cal L}_2) = b_B({\cal L}_1)
b_B({\cal L}_2)\xi_{\rm lin}(r)$, then we would have obtained exactly
the same result as Eq.~(\ref{eq:corr}).

\section{Cross correlation between dark matter annihilation and blazars}
\label{sec:Cross correlation between dark matter annihilation and
blazars}

We here derive the formulation for the angular cross correlation between
blazars and dark matter annihilation, $C_{l,BD}$, in the
subhalo-dominated case, Eqs.~(\ref{eq:cross 1-halo}) and (\ref{eq:cross
2-halo}), and in the host-halo-dominated case, Eqs.~(\ref{eq:cross
1-halo 2}) and (\ref{eq:cross 2-halo 2}), respectively in the following
subsections.
Once again, we follow the halo model approach highlighted in
Refs.~\cite{Scherrer1991,Seljak2000}.

\subsection{Subhalo-dominated case}
\label{apsub:Subhalo-dominated case}

In the halo model approach, the blazar intensity can be written as
\begin{eqnarray}
 E I_B (E,\bm{\hat n}) &=& \frac{1}{4\pi} \int d^3x \int d{\cal L}\
  \sum_i \delta_D^3(\bm x-\bm x_i)
  \nonumber\\&&{}\times 
  \delta_D({\cal L}-{\cal L}_i)
  {\cal F}_E({\cal L},z),
  \label{apeq:blazar intensity}
\end{eqnarray}
and the intensity from dark matter subhalos are Eq.~(\ref{eq:intensity
general}) with $P_\gamma$ given by Eq.~(\ref{eq:volume emissivity:
superposition}).
Therefore, the procedure of obtaining $\langle I_B I_D \rangle$ is quite
similar to that in Appendix~\ref{apsub:Dark matter substructure}.
Repeating arguments there, we arrive at Eqs.~(\ref{eq:cross 1-halo}) and
(\ref{eq:cross 2-halo}).

\subsection{Host-halo-dominated case}
\label{apsub:Host-halo-dominated case}

In the case of host-halo-dominated annihilation, the intensity is
proportional to density squared:
\begin{eqnarray}
 E I_D(E,\bm{\hat n}) &=& \int dr\ E \delta^2(r,\bm{\hat n}r)
  W([1+z]E,r)
  \nonumber\\
 &=& \int dr\ EW([1+z]E,r)
  \nonumber\\&&{}\times 
  \left[ \int d^3 x \int dM\ \sum_i \delta_D^3(\bm x-\bm  x_i)
  \right.\nonumber\\&&{}\times\left.
  \delta_D(M-M_i)
  \frac{M}{\Omega_m \rho_c} u(\bm{\hat n}r - \bm x | M) \right]^2,
  \nonumber\\
\end{eqnarray}
where $\delta (r,\bm{\hat n}r)$ is the overdensity.
Therefore, we need to evaluate the ensemble average of the product of
seed densities as follows:
\begin{widetext}
\begin{eqnarray}
\lefteqn{
\left\langle \sum_i\delta_D^3 (\bm x_1 - \bm x_i)\delta_D (M_1 -
 M_i) \sum_j\delta_D^3 (\bm x_2 - \bm x_j)\delta_D (M_2 - M_j)
 \sum_k\delta_D^3 (\bm x_3 - \bm x_k)\delta_D (M_3 - M_k)\right\rangle
}
\nonumber\\
 &=& \frac{dn}{dM_1} \frac{dn}{dM_2} \frac{dn}{dM_3}
  \left[ 1 + \xi_s^{(2)}(M_1,M_2,\bm x_1,\bm x_2)
  + \xi_s^{(2)}(M_2,M_3,\bm x_2,\bm x_3)
  + \xi_s^{(2)}(M_1,M_3,\bm x_1,\bm x_3)
  \right.\nonumber\\&&{}\left.
  + \xi_s^{(3)}(M_1,M_2,M_3,\bm x_1,\bm x_2,\bm x_3)\right]
  \nonumber\\&&{}
  + \frac{dn}{dM_1} \frac{dn}{dM_2}
  \left[ 1 + \xi_s^{(2)}(M_1,M_2,\bm x_1,\bm x_2)\right]
  \delta_D^3(\bm x_3 - \bm x_2)\delta_D(M_3-M_2)
  \nonumber\\&&{}
  + \frac{dn}{dM_2} \frac{dn}{dM_3}
  \left[ 1 + \xi_s^{(2)}(M_2,M_3,\bm x_2,\bm x_3) \right]
  \delta_D^3(\bm x_1 - \bm x_3)\delta_D(M_1-M_3)
  \nonumber\\&&{}
  + \frac{dn}{dM_1} \frac{dn}{dM_3}
  \left[ 1 + \xi_s^{(2)}(M_1,M_3,\bm x_1,\bm x_3) \right]
  \delta_D^3(\bm x_2 - \bm x_1)\delta_D(M_2-M_1)
  \nonumber\\&&{}
  + \frac{dn}{dM_1}\delta_D^3(\bm x_2 - \bm x_1)\delta_D(M_2-M_1)
  \delta_D^3(\bm x_3 - \bm x_1)\delta_D(M_3-M_1).
\end{eqnarray}
\end{widetext}
The first term represents the three-halo contribution, i.e., three
points considered $\bm x_1,\bm x_2$, and $\bm x_3$ are in three
different halos; the second to fourth terms are the two-halo
contribution, where two points are in one halo and the rest one point is
in the other halo; the last term shows the one-halo contribution, where
all three points are in the same halo.
Now, our particular focus here is that one point, say $\bm x_1$,
represents the blazar position, while the other two, $\bm x_2$ and $\bm
x_3$, the place of the dark matter annihilation.
Since the latter effect is proportional to the density squared at one
point, we have $\bm x_2 = \bm x_3$.
In this case, considering the fact that the halos are spatially
exclusive, one can omit terms except for the second (2-halo term) and
the last (1-halo term).

We then evaluate the one-halo and two-halo terms.
By substituting the above expression, we get
\begin{widetext}
\begin{eqnarray}
 \lefteqn{
  E^2 \langle I_B(E,\bm{\hat n}_1) I_D(E,\bm{\hat n}_2)
  \rangle_{1h}
  }\nonumber\\
  &=& \frac{1}{4\pi}
   \int d^3x_1 \int dr_2\ EW([1+z_2]E,r_2)\int d{\cal L}\
  \Phi({\cal L}) {\cal F}_E({\cal L},z_1)
  \left(\frac{M[{\cal L}]}{\Omega_m\rho_c}\right)^2
  u^2(\bm x_2-\bm x_1|M[{\cal L}]),\label{eq:1-halo}\\
 \lefteqn{
 E^2 \langle I_B(E,\bm{\hat n}_1) I_D(E,\bm{\hat n}_2)\rangle_{2h}
 }\nonumber\\
  &=&
  E^2 \langle I_B(E) \rangle \langle I_D(E) \rangle
  + \frac{1}{4\pi}
  \int d^3x_1 \int dr_2\ EW([1+z_2]E,r_2)\int d{\cal L}\
  \Phi({\cal L}) {\cal F}_E({\cal L},z_1)
  \int dM_1 \frac{dn(M_1)}{dM}
  \left(\frac{M_1}{\Omega_m\rho_c}\right)^2
  \nonumber\\&&{}\times
  \int d^3y_1\ u^2(\bm x_2-\bm y_1|M_1)
  \xi_s^{(2)}(M[{\cal L}],M_1,\bm x_1,\bm y_1),\label{eq:2-halo}
\end{eqnarray}
\end{widetext}
where we used Eq.~(6) of AK06 to reach $\langle I_B(E) \rangle \langle
I_D(E) \rangle$ in the second expression.
We again use small separation approximation, $d^3x_1 dr_2 / 4\pi =
r_1^2 dr_1dr_2 = r^2 dr d\eta$, and get
\begin{eqnarray}
 \lefteqn{
  E^2 \langle I_B(E,\bm{\hat n}_1) I_D(E,\bm{\hat n}_2)\rangle
  }\nonumber\\
  &=&
  \int dr d\eta\ r^2 EW([1+z]E,r) \int d{\cal L}\
  \Phi({\cal L}) {\cal F}_E({\cal L},z)
  \nonumber\\&&{}\times
  \left[\left(\frac{M[{\cal L}]}{\Omega_m\rho_c}\right)^2
   u^2(\eta\bm{\hat r} + r\theta\bm{\hat\theta}|M[{\cal L}])
   \right.\nonumber\\&&{}\left.+
  \int dM^\prime\ \frac{dn(M^\prime)}{dM}
  \left(\frac{M^\prime}{\Omega_m\rho_c}\right)^2
  \int d^3y\
  \right.\nonumber\\&&{}\left.\times
  u^2(\bm x_2 - \bm y|M^\prime)
  \xi_s^{(2)}(M[{\cal L}],M^\prime,\bm x_1,\bm y)\right].
\end{eqnarray}
To obtain the angular power spectrum, we Fourier-transform this
expression.
Using the similar relations to Eq.~(\ref{apeq:fourier}), we finally arrive
at Eqs.~(\ref{eq:cross 1-halo 2}) and (\ref{eq:cross 2-halo 2}).


\end{document}